\documentclass[prb,twocolumn,showpacs]{revtex4}
\usepackage{graphicx}
\usepackage{amsmath}
\usepackage{amssymb}
\usepackage{dcolumn}
\usepackage{float}
\usepackage{bm}
\usepackage{color,soul}
\usepackage[breaklinks=true,colorlinks,citecolor=blue,linkcolor=blue,urlcolor=blue]{hyperref}

\newcommand{\nn}{\nonumber}

\DeclareMathAlphabet{\bi}{OML}{cmm}{b}{it}

\def\be{\begin{equation}}
\def\ee{\end{equation}}
\def\bearr{\begin{eqnarray}}
\def\eearr{\end{eqnarray}}

\def\bs{\boldsymbol}

\begin{document}
\title{Phonon-drag magnetoquantum oscillations in graphene}
\bigskip

\author{S. S. Kubakaddi}
\email{sskubakaddi@gmail.com}
\author{Tutul Biswas}
\email{tbtutulm53@gmail.com}

\author{Tarun Kanti Ghosh}
\email{tkghosh@iitk.ac.in}
\normalsize
\affiliation
{$\color{blue}{^\ast}$ Department of Physics, K. L. E. Technological University, Hubballi-580 031, Karnataka, India\\
$\color{blue}{^\dagger}$ Department of Physics, Vivekananda Mahavidyalaya, Burdwan-713 103, West Bengal, India\\
$\color{blue}{^\ddagger}$ Department of Physics, Indian Institute of Technology-Kanpur,
Kanpur-208 016, Uttar Pradesh, India}
\date{\today}

\begin{abstract}
A theory of low-temperature phonon-drag magnetothermopower $S_{xx}^g$ is 
presented in graphene in a quantizing magnetic field. $S_{xx}^g$ is found 
to exhibit quantum oscillations as a function  of magnetic field $B$ and 
electron concentration $n_e$. Amplitude of the oscillations is found to 
increase (decrease) with increasing $B$ ($n_e$). The behavior of $S_{xx}^g$ is also
investigated as a function of temperature. A large value of $S_{xx}^g $
($\sim$ few hundreds of $\mu $V/K) is predicted.  Numerical values of 
$S_{xx}^g $ are compared with the measured magnetothermopower $S_{xx}$ and
the diffusion component $S_{xx}^d$ from the modified Girvin-Jonson theory.  
\end{abstract}


\maketitle

\section{Introduction}
Graphene, a monolayer of carbon atoms arranged in a honeycomb lattice 
of hexagons, has a unique band structure. Its electronic states, at 
the points ${\bf K}$ and ${\bf K}^{\prime}$ of the Brillouin zone, 
have a linear dispersion relation, described by the Dirac equation.
It is an ambipolar material with zero effective mass of the carriers 
and zero energy gap. Its electrical transport properties have been 
studied extensively \cite{neto,sarma}, since its discovery \cite{novo,novo1,kim}, 
with host of intriguing 
phenomena due to its unusual band structure. In a quantizing magnetic 
field ${\bf B}$, graphene exhibits the integer quantum Hall effect (QHE) \cite{novo1,kim}, 
with novel features different from those in conventional two-dimensional 
electron gas (2DEG), particularly at $n=0$ Landau level (LL). 

Thermopower ${\bf S}$, an electric field ${\bf E}$ generated in a sample
due to unit temperature gradient i.e. ${\bf S} = {\bf E}/(-{\bs \nabla} T) $,
has been another powerful tool for probing carrier transport. 
Application of magnetic field ${\bf B}$, in addition to a temperature 
gradient ${\bs \nabla} T$, provides valuable experimental tool to investigate 
magnetothermoelectric effects. In a 2DEG, in the $xy$ plane, a temperature
gradient ${\bs \nabla} T \parallel x$-axis and magnetic field 
${\bs B}\parallel z$-axis generates electric field ${\bf E}$, in the sample, with components
$E_x = S_{xx} (-{\bs \nabla} T)_x$ and $E_y = S_{yx} (-{\bs \nabla} T)_x$, where
the thermopower $S_{xx}$ and the Nernst-Ettingshausen coefficient $S_{yx}$ are 
the tensor components of ${\bf S}$. Quantization effects due to magnetic field 
are reflected in thermopower.

In conventional 2DEG of GaAs heterojunctions (HJs) and Si-MOSFETs,
magnetothermopower is investigated in detail, experimentally and theoretically, in
the quantum Hall regime \cite{but,flet,flet-2003,tsa}. Measured magnetothermopower
tensor components $S_{xx}$ and
$S_{yx}$ exhibited oscillations as a function of magnetic field, arising due to 
crossing of Landau level by the Fermi level due to either change in carrier
concentration or magnetic field. There are two additive and independent contributions
to thermopower ${\bf S}$. In a ${\bs \nabla} T$, diffusion component ${\bf S}^{ d}$ arises
due to diffusion of carriers and the phonon drag component ${\bf S}^{g}$ arises
due to the non-equilibrium phonons transferring some of their momentum
to the electrons via electron-phonon (el-ph) scattering. The oscillatory 
behavior of diffusion component is explained by the theory of Jonson and Girvin \cite{jonson} and
Oji \cite{,oji}.
It is established that for about $ 0.1 <T<10 $ K, the $S^g$ contribution dominates
$S$ in GaAs HJs\cite{but,flet,flet-2003}.
$ S^{ g}$ is important because it gives directly el-ph coupling
and is independent of impurity scattering, unlike mobility. 

The study of phonon-drag magnetothermopower $S^{ g}$, in conventional 2DEG,
began with pioneering experimental work of Fletcher et al \cite{flet-1986} showing 
the quantum oscillations as a function of $B$. Its $B$ and $T$ dependence were explained by 
developing the theory of $S^{g}$ \cite{kuba,lyo,from}, by modifying the Boltzmann 
theory of phonon-drag in bulk semiconductors \cite{ger,puri}, following the $\Pi$-approach 
due to Herring \cite{herring}. 

In graphene, the experimental and theoretical investigations of thermoelectric
effects in zero and quantizing magnetic field are being intensively pursued \cite{sank}. 
In monolayer graphene, experimental data of $S$ vs $T$, in zero magnetic field,
in temperature regime 10-300 K show largely linear behavior suggesting that 
the mechanism for thermopower is diffusive \cite{zuev}.
Thermopower measurements in quantum Hall regime are carried out as a function
of magnetic field for different gate voltage (i.e. for different carrier
concentration) and temperatures \cite{zuev,check,wei,wu}. Magnetothermopower $S_{xx}$ and 
Nernst-Ettingshausen coefficient $S_{yx}$ have shown the oscillatory behavior as
a function of magnetic field. The behavior of $S_{xx} $ and $S_{yx}$ are
in agreement with the generalized Mott relation, extending the theory of Jonson and
Girvin \cite{jonson} to graphene \cite{zuev}.
The peak values of $S_{xx}$ are predicted to be given by 
$S_{xx}^{\rm peak}=(-k_B/e)\ln2/n $, noting that Jonson-Girvin theory fails for Landau
level $n=0$. Similar observations are made in high mobility samples of 
graphene \cite{wu}. Zero-field  non-linear $T$ dependence is attributed to the 
screening \cite{das-sarma}.  Measured strong 
quantum oscillations as a function of $B$ are understood by evolution of the 
density of states at the Fermi level and $S_{xx}$ becoming zero when the Fermi level
lies in the localized states, because of absence of diffusion \cite{wu}. In all these 
measurements, it is observed that phonon-drag thermopower component is
absent and no evidence of phonon-drag magnetoquantum oscillations, even at 
low temperatures, attributing to the weak el-ph coupling. 
We believe that, equally important reason for the absence of phonon-drag 
component in these samples may be due to their small size
($\sim 300$ nm). It is about $10^{4}$ times smaller than the samples of GaAs HJs
($\sim $ mm in which $S_{xx}$ is large and about $\sim $ mV/K) \cite{but}. At low temperatures,
the smaller dimension of the sample sets the limit for phonon mean free path
$\Lambda$, in the boundary scattering regime, as $S^g \sim \Lambda$. We 
expect the phonon-drag to be significant in large samples (few $\mu$m) for e.g.
in the samples of Nika et al \cite{nika}. Moreover, to know the significant 
contribution of phonon-drag contribution, more data of $S$ is required at 
low temperature covering sub-Kelvin region in pure samples.

The theory \cite{kuba-2009} of zero magnetic field $S^g$ has been 
developed in monolayer graphene, in the boundary scattering regime, as a function 
of temperature $T (\leq 10$ K) and electron concentration $n_e$ for the phonon mean free
path $\Lambda \sim 10 \mu$m (closer to the samples of Nika et al \cite{nika}).
At about $10$ K, $S^g \sim $ 10  $\mu$V/K. 
This value is nearly same order
of magnitude as that of predicted $S^{d}$ with the peak values few tens of 
$\mu$V/K, by the modified Jonson-Girvin formula. We have to note that,
unlike $S^g$, the latter is independent of sample size. 

It would be interesting to study the effects
of magnetic field quantization on phonon-drag thermopower $S_{xx}^g$.
In the present work, we theoretically investigate the  phonon-drag magnetothermopower
as a function of magnetic field $B$, electron concentration $n_e$,  and temperature $T$.
We explore the circumstances and possibilities of its significant contribution
to the measured magnetothermopower, by tuning the parameters $B$, $n_e$, $T$, and $\Lambda$.
For comparison, we also compute diffusion component $S_{xx}^d$.
The qualitative comparison of our calculations is made with the  experimental observations.

This paper is organized as follows. In Sec. II, we provide formalism of
phonon-drag thermopower in presence of quantizing magnetic field.
In Sec. III, we present our results and discussion. 
A summary of our work is provided in Sec. IV.

\section{Formalism of phonon-drag magnetothermopower}
In the following we proceed with the calculations by appropriately modifying the
theory of Fromhold et al \cite{from} for the monolayer graphene. We consider an isotropic
and homogeneous 2DEG of graphene in the $xy$-plane with the magnetic field
${\bf B} = (0,0,B)$ along the $z$-direction. In presence of an electric field 
${\bf E} $ (along $x$-axis) electrons are assumed to be accelerated
isothermally (${\bs \nabla} T =0$). In the steady state, the non-equilibrium
distribution of the electrons in state 
$\alpha$ is given by $ f_{\alpha} = f_{\alpha}^0 + f_{\alpha}^1$, where
$f_{\alpha}^0 = [\exp\{(E_{\alpha}^0-\mu_F)/k_BT\} +1 ]^{-1}$ is the thermal
equilibrium distribution function in absence of the electric field for state
$\alpha$, $\mu_F$ is the chemical potential and $f_{\alpha}^1$ is the first-order
perturbation due to electric field $E$. These non-equilibrium electrons transfer
some of their momentum to the 2D phonons through the el-ph coupling. This causes
perturbation in the phonon distribution which is given by 
$N_{\bf q} = N_{\bf q}^0 + N_{\bf q}^1 $, where 
$ N_{\bf q}^0 = [\exp(\hbar \omega_{\bf q}/k_BT) - 1]^{-1} $ is the equilibrium
distribution of the phonons of energy $\hbar \omega_{\bf q}$ and the wave vector
${\bf q}$. Here, $N_{\bf q}^1 $ is the perturbation in the phonon distribution, 
due to electric field, producing the heat current density ${\bf U}$.

We confine our attention to the linear transport regime at liquid helium
temperature. Then it is necessary to consider only acoustic phonons, with the 
2D character, which interact weakly with the 2D electrons.
The phonon heat current density, noting that $N_{\bf q}^0 $ will not contribute,
is given by 
\begin{eqnarray} \label{heat}
{\bf U} = A_0^{-1} \sum_{\bf q} N_{\bf q}^1 \hbar \omega_{\bf q} v_{\bf q},
\end{eqnarray}
where $A_0 = L_x L_y$ is the area of graphene sample and $v_{\bf q} $ is the phonon 
group velocity.

In the linear response regime ($ N_{\bf q}^1 \propto E$), the heat current 
density is given by ${\bf U} = {\bf M E} = {\bs \Pi}{\bf J}$, where ${\bf M}$ is 
the thermoelectric tensor, ${\bs \Pi} = {\bf M}/{\bs \sigma} $ is the Peltier 
coefficient tensor and  ${\bs \sigma}$ is the electrical conductivity tensor. 
From the Onsager relation, the thermopower tensor is defined as ${\bf S} = {\bs \Pi}/T$. 
Using the Onsager symmetry relations it has been shown that \cite{from} 
$ T S_{xx} = \rho_{xx} M_{xx} - \rho_{yx} M_{yx} $ and
$ T S_{yx} = \rho_{yx} M_{xx} + \rho_{xx} M_{yx} $, where $ S_{xx} $ and $S_{yx} $
are, respectively, thermopower and Nernst-Ettinshausen coefficient, $M_{xx} $ and
$M_{yx} $ are the components of tensor ${\bf M} $ and $\rho_{xx} $ and
$\rho_{yx} $ are the components of electrical resistivity tensor ${\bs \rho}$.
Hence, the calculation of ${\bf U}$ will facilitate the calculation of ${\bf M} $ and
hence $S_{xx}$ and $S_{yx} $.

The solution for $N_{\bf q}^1$, in the linear response regime, is found to be
\begin{eqnarray} \label{phonon}
N_{\bf q}^1 = \frac{g}{k_BT} \sum_{\alpha,\alpha^{\prime}} \tau_q
\Gamma_{\alpha,\alpha^{\prime}}({\bf q}) 
\Big[\frac{f_{\alpha}^1}{(\partial f_{\alpha}^0/\partial E_{\alpha})}
- \frac{f_{\alpha^{\prime}}^1}
{(\partial f_{\alpha^{\prime}}^0/\partial E_{\alpha^{\prime}})}\Big],
\end{eqnarray}
where $g = g_s g_v$, $g_s(g_v) $ is the spin (valley) degeneracy,
and 
\begin{eqnarray}\label{phonon-a}
& & \Gamma_{\alpha,\alpha^{\prime}}({\bf q}) = P_{\alpha,\alpha^{\prime}}^{{\rm ab}0}({\bf q})
f_{\alpha}^0(1- f_{\alpha^{\prime}}^0), \\
& & P_{\alpha,\alpha^{\prime}}^{{\rm ab}0}({\bf q})  =  \frac{2\pi}{\hbar} 
|C_{\alpha,\alpha^{\prime}}({\bf q})|^2 N_{\bf q}^0 
\delta(E_{\alpha^{\prime}} - E_{\alpha} - \hbar \omega_{\bf q}). \label{phonon-b}
\end{eqnarray}
Here, $P_{\alpha,\alpha^{\prime}}^{{\rm ab}0}({\bf q})$ 
is the transition probability, in equilibrium, for the electron scattering 
from state $\alpha$ to state $\alpha^{\prime}$ by absorbing a phonon and
$ |C_{\alpha,\alpha^{\prime}}({\bf q})|^2 $ is the square of the electron-acoustic 
phonon interaction matrix element. 
Equation (2) is the general to the extent that, it is independent of the electronic
structure and the type of el-ph coupling.

In a quantizing magnetic field ${\bf B} = (0,0,B) $ with the Landau gauge
${\bf A} = (0, Bx,0) $, the eigenfunctions and  energy eigen values are given 
in Ref. [\onlinecite{matu}]. 
The energy eigen values are $E_{\alpha}^0 = E_{n,k_y}^0 = \hbar \omega_c \sqrt{2n} $,
where $\alpha \equiv (n,k_y), n=0,1,2...$  is the Landau level quantum number, 
$k_y$ is the electron wave vector in the $y$-direction, 
$\omega_c = v_F/l_0 $ is the cyclotron frequency, $v_F = 1 \times 10^6 $ m/s is the  Fermi velocity of 
electron in graphene, and $l_0 = \sqrt{\hbar/(eB)}$ is the magnetic length. The
el-ph matrix element is given by (see Appendix A for details)
\begin{eqnarray}\label{MatEl}
\vert C_{\alpha,\alpha^{\prime}}({\bf q})\vert^2  = \vert C({\bf q})\vert^2 \vert J_{n,n^{\prime}}(u)\vert^2 
\delta_{k_{y}^{\prime}, k_y+q_y},
\end{eqnarray}
where $\vert C({\bf q})\vert^2 $ is the matrix which describes the el-ph coupling strength and
$ \vert J_{n,n^{\prime}}(u)\vert $ with $ u = q^2 l_0^2/2$, is the matrix element describing 
the scattering between Landau levels.

In presence of crossed electric field ${\bf E} = (E,0,0)$ and magnetic field ${\bf B}=(0,0,B)$, 
the energy spectrum of graphene can be found exactly \cite{baskaran,peres}.
In the linear response regime, where the applied electric field
is low enough, one can obtain energy eigen value for the magnetic state $\alpha$, approximately, as 
$E_{\alpha} \simeq E_{\alpha}^0 + e E x_{\alpha}$ 
by expanding the exact expression given in Refs. [\onlinecite{baskaran,peres}] up to 
first-order in $E/(v_FB)$. Here, $x_{\alpha} = k_y l_0^2$. This
is nothing but the first-order energy correction due to the week external electric field.
Further, assuming that the form of distribution function retains the same with the
modified energy, we expand $f_{\alpha} = f_0(E_{\alpha}) = f_0 (E_{\alpha}^{0})
- eE x_{\alpha} [\partial f(E_{\alpha}^{0})/\partial E_{\alpha}^{0}]
( = f_0(E_{\alpha}^{0}) + f_{\alpha}^1) $, which gives 
$f_{\alpha}^1/[\partial f(E_{\alpha}^{0})/\partial E_{\alpha}^{0}] = - e E x_{\alpha} $.
Then, Eq. (\ref{phonon}), using the momentum conservation 
$ k_{y}^{\prime}= k_y+q_y $, for the chosen Landau gauge gives,
\begin{eqnarray} \label{phonon1}
N_q^1 = \frac{g  e E l_0^2}{k_B T} 
\sum_{\alpha \alpha^{\prime}} \tau_q \Gamma_{\alpha \alpha^{\prime}}({\bf q}) q_y.
\end{eqnarray}
 
In order to make $ N_q^1 $ linear in $E$, we set all terms in 
$ \Gamma_{\alpha \alpha^{\prime}}(q)$ independent of $E$. Inserting Eq. (\ref{phonon1})
into Eq. (\ref{heat}) we write $ {\bf U} ={\bf ME }$ and take the phonon group velocity 
components $ v_{q}^x = (q_x/q) v_s $ and $ v_{q}^y = (q_y/q) v_s $, $v_s$ 
being the acoustic phonon velocity in graphene.
Then, the two components of the thermoelectric tensor ${\bf M}$ are given by
\begin{eqnarray}
& & M_{xx} = \frac{g e l_0^2 v_s}{A_0 k_B T} 
\sum_{\bf q} \tau_{ q} \Gamma({\bf q})(q_xq_y/q) \hbar \omega_{\bf q}\\
& & M_{yx} = \frac{g e l_0^2 v_s}{A_0 k_B T}
\sum_{\bf q} \tau_{ q} \Gamma({\bf q})(q_y^2/q) \hbar \omega_{\bf q} \label{mxy},
\end{eqnarray}
where 
\begin{eqnarray} \label{gamma}
\Gamma({\bf q}) = \sum_{\alpha \alpha^{\prime}} \Gamma_{\alpha \alpha^{\prime}}({\bf q}).
\end{eqnarray}

When we carry out the angular integration, it can be seen that $M_{xx} =0$ because
of the $xy$ isotropy. However, this is shown as the limitation of this theory as the 
experimental results of $M_{xx}$ in conventional 2DEG show its non-zero value \cite{from}. 
Later calculations in these systems, taking into account of 
anisotropy of electrons and phonons, remove
this limitation \cite{but-1998}. However, in the present work we undertake the evaluation of only
$M_{yx} $, as we have considered $xy$ isotropy of the system.

Using Eqs. (\ref{phonon-a}) and (\ref{phonon-b}), Eq. (\ref{gamma}) turns out to be 
\begin{eqnarray}\label{Gamma10}
\Gamma({\bf q}) & = & \frac{2\pi}{\hbar}\sum_{\alpha} \sum_{\alpha^{\prime}} 
\vert C_{\alpha,\alpha^{\prime}}({\bf q}) \vert^2 N_{\bf q}^{0} 
f^0(E_{\alpha}^0) \nn \\ 
& \times & [1 - f^0(E_{\alpha^{\prime}}^0)] 
\delta(E_{\alpha^{\prime}}^0 - E_{\alpha}^0 - \hbar \omega_{\bf q}).
\end{eqnarray}

Summation of over $k_{y}^{\prime} $ is carried out replacing it by
$k_y + q_y$. Since the integrand is independent of $k_y$, summation over
$k_y$ simply gives $A_0/2\pi l_0^2$.

In the presence of disorder, the energy levels (in zero electric field)
$E_{n,k_y}^0 $ and $ E_{n,k_y+q_y}^0 $ are randomized by LL broadening. 
A simple system average is taken by integrating over $\epsilon = E_{n,k_y}^0 $
and  $\epsilon^{\prime} = E_{n,k_y+q_y}^0 $ with the weight factor 
$\rho(\epsilon - E_{n}^0) \rho(\epsilon^{\prime} - E_{n^{\prime}}^0) $, 
where $E_{n}^0 (E_{n^{\prime}}^0) $ is the energy of the $n (n^{\prime})$-th
LL in absence of disorder and $\rho(x)$ is the LL density of states
with convenient line shape function.
Now Eq. (\ref{Gamma10}) becomes
\begin{eqnarray}
\Gamma({\bf q}) & = & \frac{A_0}{2\pi l_0^2}\frac{2\pi}{\hbar} \sum_n \sum_{n^\prime} 
\vert C_{n,n^\prime}({\bf q})\vert^2 N_{\bf q}^0 \nn \\
& \times & \int d\epsilon \, d\epsilon^\prime
f^0(\epsilon) [1- f^0(\epsilon^{\prime})] \nn  \\ 
& & \delta(\epsilon^\prime - \epsilon - \hbar \omega_{\bf q}]) 
\rho(\epsilon - E_{n}^0) \rho(\epsilon^{\prime} - E_{n^{\prime}}^0).
\end{eqnarray}

Integration with respect to $\epsilon^{\prime} $, using the Dirac delta function,
gives
\begin{eqnarray} \label{gamma1}
\Gamma({\bf q}) = \frac{A_0}{\hbar l_{0}^2} \sum_n \sum_{n^\prime} 
\vert C({\bf q})\vert^2 \vert J_{n,n^\prime}(u) \vert^2 N_{\bf q}^0 
I_{n n^\prime}(\hbar \omega_{\bf q}),
\end{eqnarray}
where
\begin{equation}
I_{nn^\prime}(\hbar \omega_{\bf q}) = \int d\epsilon 
f^0(\epsilon) [1- f^0(\epsilon + \hbar \omega_{\bf q} )] 
\rho(\epsilon - E_{n}^0) \rho(\epsilon + \hbar \omega_{\bf q} - E_{n^{\prime}}^0).
\end{equation}

Since the phonon-drag thermopower is important at low temperature, 
the energy of acoustic phonons involved is so small that only intra LL scattering
is possible. Thus we set $ n=n^\prime$ in Eq. (\ref{gamma1}). Then, with 
\begin{equation}\label{JnnS}
\vert J_{nn}(u)\vert^2 = \frac{e^{-u}}{4} \Big[ L_n(u) + L_{n-1}(u)\Big]^2,
\end{equation}
Equation (\ref{gamma1}) gives
\begin{equation} \label{gamma2}
\Gamma({\bf q}) = \frac{A_0}{\hbar l_{0}^2} \vert C({\bf q})\vert^2 N_{\bf q}^0 
\sum_n \vert J_{nn}(u)\vert^2 I_{nn}(\hbar \omega_{\bf q}).
\end{equation}
Intra LL transitions are possible as the energy levels are broadened.

Using Eq. (\ref{gamma2}) in Eq. (\ref{mxy}), we get
\begin{eqnarray}
M_{yx} & = & \frac{g e v_s}{\hbar k_BT} \sum_{\bf q} \tau_{q} (q_y^2/q) 
\hbar \omega_{\bf q} \vert C({\bf q})\vert^2 N_{\bf q}^0 \nn \\ 
& \times & \sum_n \vert J_{nn}(u)\vert^2 I_{nn}(\hbar \omega_{\bf q}).
\end{eqnarray}
The summation over ${\bf q}$ is converted into integration as
\begin{equation}
\sum_{\bf q} \rightarrow \frac{A_0}{(2\pi)^2} \int_0^{\infty} q dq 
\int_0^{2\pi} d\theta.
\end{equation}
Angular integration coming through $ q_y = q \sin \theta $ gives $\pi$.
Then
\begin{eqnarray}
M_{yx} & = & \frac{g e A_0}{4 \pi \hbar^4 v_s^2 k_BT} \int_0^{\infty}
d (\hbar \omega_{ q}) \tau_q (\hbar \omega_{ q})^3
\vert C(q)\vert^2 N_{q}^0 \nn \\
& \times & \sum_n \vert J_{nn}(u)\vert^2 I_{nn}(\hbar \omega_{q}).
\end{eqnarray}
Substituting for $\vert C(q)\vert^2 = D^2 \hbar \omega_q/(2\rho_m A_0 v_s^2)$,
where $D$ is the acoustic phonon deformation potential coupling constant and
$\rho_m$ is the areal mass density of graphene, we obtain
\begin{eqnarray}
M_{yx} & = & \frac{g e D^2}{8 \pi \rho_m \hbar^4 v_s^4 k_BT} \int_0^{\infty}
d (\hbar \omega_{q}) \tau_q (\hbar \omega_{q})^4
 N_{q}^0 \nn \\
& \times & \sum_n \vert J_{nn}(u)\vert^2 I_{nn}(\hbar \omega_{q}).
\end{eqnarray}

From Onsager relation, with $ M_{xx} =0 $, 
we have $ T S_{xx}^g = - \rho_{yx} M_{yx}$. Taking $\rho_{yx} = B/(n_e e)$\cite{tiwari}, where
$n_e$ is the electron density and expressing $B$ in terms of $l_0^2$, 
we get
\begin{eqnarray}
S_{xx}^g & = &-\frac{gk_B D^2}{8 \pi e \rho_m \hbar^3 v_s^4 l_0^2 n_e (k_BT)^2} 
\int_0^{\infty} d (\hbar \omega_{\bf q}) \tau_q (\hbar \omega_{\bf q})^4
 N_{q}^0 \nn \\
& \times &
\sum_n \vert J_{nn}(u)\vert^2 I_{nn}(\hbar \omega_{ q}).
\end{eqnarray}

We note that this equation can also be obtained by following the method of 
Kubakaddi et al \cite{kuba} for conventional 2DEG ignoring $\Gamma(q)$ compared 
to $ 1/\tau_q$.

\section{Results and Discussion}
Since $S_{xx}^g \sim \Lambda$ and $D^2$, it is essential to choose the reasonable 
values of these parameters. We numerically evaluate $S_{xx}^g$, for $T\leq20$ K 
in the boundary scattering regime for which $\tau_q = \Lambda/v_s$, where $\Lambda$ is 
the phonon mean free path. Normally, $\Lambda$ is taken to be the smaller
dimension of the sample. Thermal conductivity calculations are demonstrated with 
$\Lambda$ chosen in the range of $3$-$30$ $\mu$m and the choice of $\Lambda=5$ $\mu$m 
is giving reasonable agreement with the measured thermal conductivity \cite{nika,nika2}.
Nika et al\cite{nika}, to  fit the thermal conductivity data, use the effective phonon mean
free path $\Lambda_{\rm eff} = \Lambda(1+p)/(1-p)$ by modulating the smallest dimension of the 
sample using specular parameter $p =0.9$, which enhances $\Lambda$ by a factor of $20$. 
The value of $0\leq p\leq 1$ is determined by the
roughness of the graphene edges. 
To present our calculations  we choose a reasonable  value of  $\Lambda= 10$ $\mu$m. 

In the literature there is a range of $D=3$-$30$ eV.  We chose $D=20$ eV which is  
closer to the values of $D$, for unscreened el-ph interaction, used to fit the experimental data of some of the 
transport properties\cite{baker,huang,bist,silva}. The line shape function $\rho(x)$
of LL is taken to be Lorentzian with the width $\Gamma = C\sqrt{B}$, 
where $C = 0.5$ meV/$\sqrt{\rm Tesla}$. 
Other parameter values used are: $\rho_m = 7.6 \times 10^{-7}$ Kg/m$^2$ and 
$v_s=2\times 10^4$ m/s.

\begin{figure}[h!]
\begin{center}\leavevmode
\includegraphics[width=100mm,height=60mm]{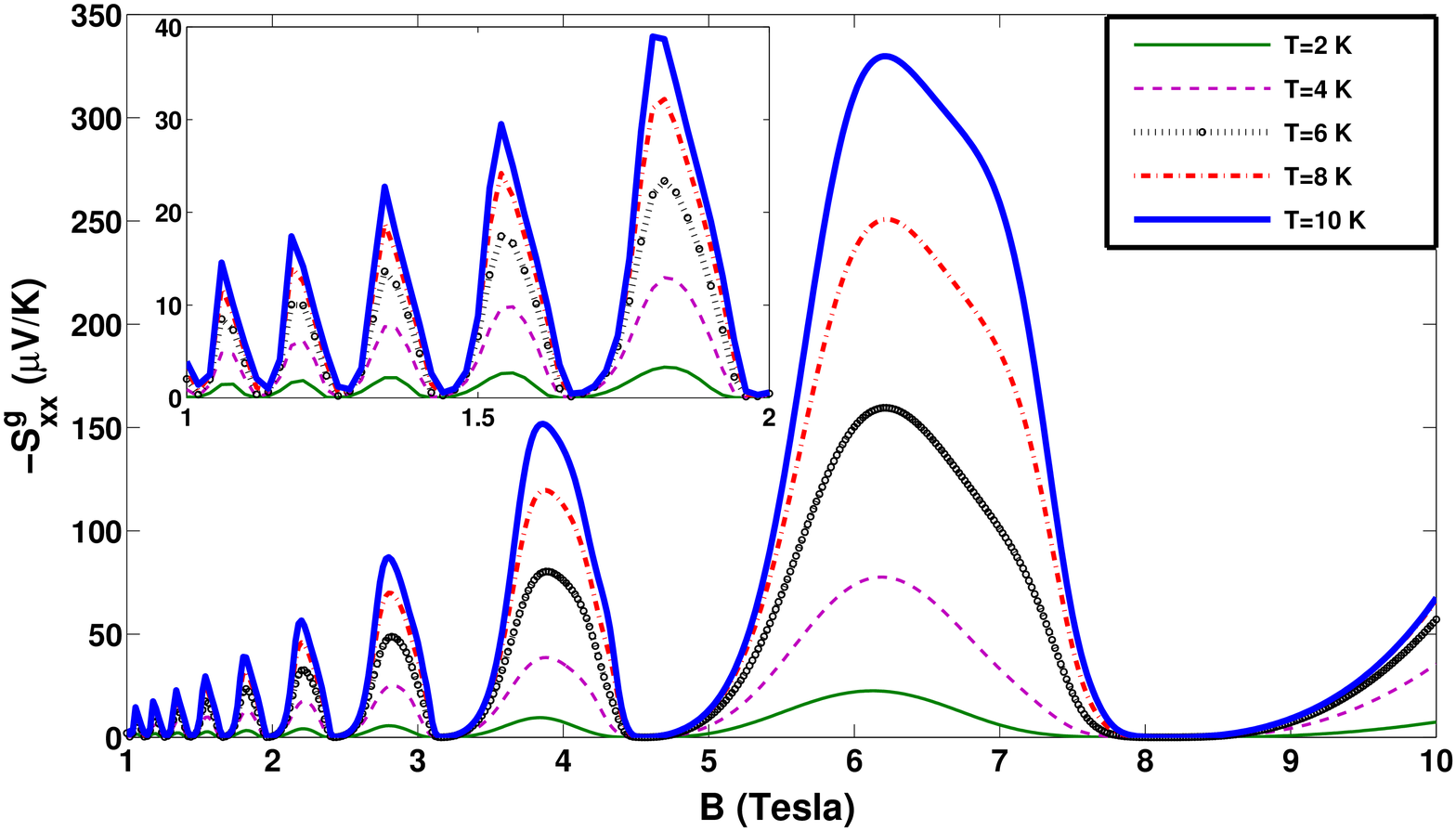}
\caption{(Color online) Plots of phonon-drag thermopower $S_{xx}^g $ versus magnetic 
field $B$ at five different temperatures, namely, $T=2, 4, 6, 8,$ and $10$ K
for $n_e=10^{16}$ m$^{-2}$ and $ \Lambda = 10$ $\mu$m. 
Inset shows the behavior of $S_{xx}^g$ in the low field regime.}
\label{Fig1}
\end{center}
\end{figure}

In Fig. 1, $S_{xx}^g$ is shown as a function of $B$, for $T=2,4,6,8,10$ K for 
$n_e=1\times10^{16}$ m$^{-2}$. We see that $S_{xx}^g$ is oscillatory with the height 
of the peak increasing with the increasing $B$. The position of the peak occurs when 
the Fermi energy is in the localized state of LL.
Interestingly, our calculations show large  peak values of the order of 
few hundreds of  $\mu$V/K which is closer to the values observed in GaAs HJs \cite{flet-1986,tie}.
We would like to point out that the sample size in GaAs HJs ($\sim$ few mm) is about
two orders of magnitude larger than the size of the graphene sample chosen here.
The size of the graphene sample ($300$ nm) in the experiment of Zuev et al \cite{zuev}
is about $30$ times smaller than the value of $\Lambda$ used in the present calculation.
Scaling the $\Lambda$ down by $30$ times, we get the peak values of  $S_{xx}^g$ few 
tens of $\mu$V/K which is comparable to the measured values.

Dependence of  $S_{xx}^g$ on $n_e$ is shown in Fig. 2 for three 
different magnetic fields, namely $B=2.82, 3.88$, and $6.20$ Tesla, at $T=5$ K 
taking $\Lambda=10$ $\mu$m. Again, the behavior is found to be oscillatory. This 
is similar to the behavior observed (as function of gate voltage) in the
experiment of Zuev et al\cite{zuev}. The peak value of $S_{xx}^g$ is decreasing 
with the increasing $n_e$. This is similar to the $n_e$ dependence of zero field
$S^g$ and $S^{d}$\cite{kuba-2009}. Also, it is found that the peak values are 
smaller for smaller $B$. 
The number of oscillations contained in $S_{xx}^g$ gets reduced with the increase 
of magnetic field. This is due to the increase of the separation between LLs with 
increasing magnetic field.  
Interestingly, we note that the position of the peak
values corresponding to three different $B$ are coinciding at 
$n_e=(1,4,7$ and $10) \times 10^{16}$ m$^{-2}$.

\begin{figure}[h!]
\begin{center}\leavevmode
\includegraphics[width=100mm,height=60mm]{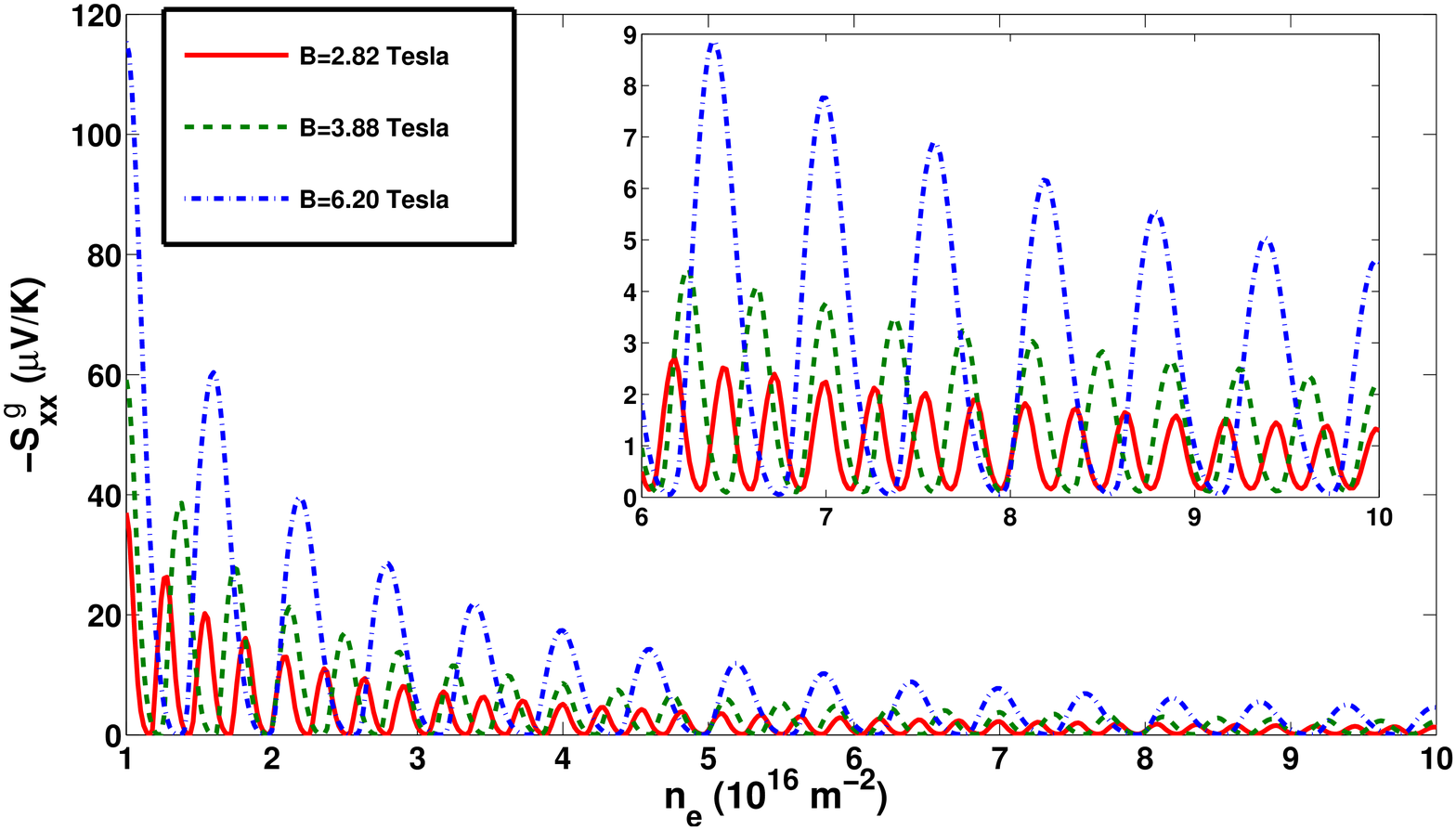}
\caption{(Color online) Plots of phonon-drag thermopower  $S_{xx}^g $ 
versus carrier density $n_e$ at different magnetic fields, namely, 
$B=2.82, 3.88,$ and $6.20$ Tesla for $T=5$ K and $\Lambda = 10$ $ \mu$m.
The behavior of $S_{xx}^g$ around the higher density regime 
is depicted in the inset.}
\label{Fig1}
\end{center}
\end{figure}

In Fig. 3(a), we have shown $S_{xx}^g$ as a function of 
$n_e$ for $\Lambda=300$ nm (as taken in Ref. [\onlinecite{zuev}])
at temperatures $T= 4.2, 10,$ and $20$ K for a magnetic field $B=8.8$ Tesla. 
For comparison, we have calculated diffusion component $S_{xx}^d$ as a function 
of $n_e$, for the same $T$ and $B$, using modified Girvin-Jonson theory 
(see Ref. [\onlinecite{check}]) 
and it is shown in Fig. 3(b). Note that $S_{xx} ^{\rm d}$ is also found to decrease
with increasing $n_e$. According to the Girvin-Jonson theory,
the peak values due to diffusion component are  given by $(k_B/e) {\rm ln}2/n$. 
$S_{xx}^d$ is found to be much greater than $S_{xx}^g$. For 
example, for $n_e \sim 1.5 \times 10^{12}$ cm$^{-2}$, at 4.2 K (10 K) 
$S_{xx}^d$ is nearly ten (three) times greater than $S_{xx}^g$. 
The total $S_{xx}=S_{xx}^d+S_{xx}^g$ is shown as a
function of $n_e$ in Fig. 3(c) and it is increasing with $T$.

We would like to point out that in graphene,
the peak values of diffusion thermopower $S_{xx} ^{d}$ are quantized as $(k_B/e){\rm ln}2/n$ which
differs from the peak value quantization $(k_B/e){\rm ln}2/(n +1/2)$ corresponding to the conventional 2DEG.
This difference is attributed to the existence of a non-trivial Berry phase $\pi$ in graphene\cite{check}.
Unlike diffusion thermopower, it is difficult to establish such
peak value quantization for $S_{xx}^g$. However, a careful observation of Fig. 3(a) \& 3(b) dictates us
that $S_{xx}^g$ follow $S_{xx}^d$ with respect to the locations of peaks.
Moreover, the locations of thermopower peaks in 2DEG and graphene are expected to be different due to different 
Landau level structures as was found in the case of conductivity oscillations\cite{matu}.

\begin{figure}[h!]
\begin{center}\leavevmode
\includegraphics[width=115mm,height=65mm]{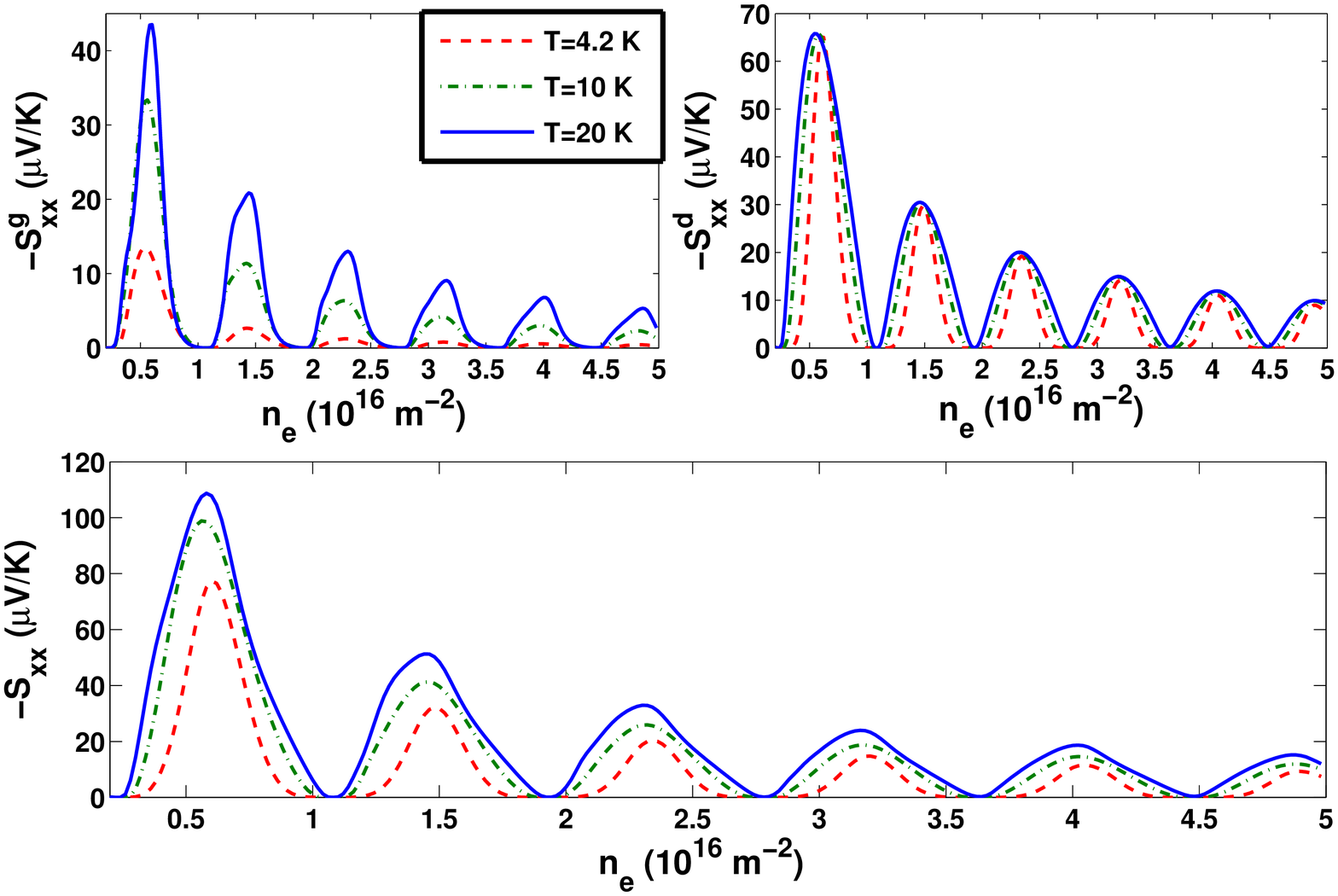}
\caption{(Color online) Plots of thermopower versus carrier density $n_e$ 
at different temperature for $B=8.8$ Tesla and $\Lambda=300$ nm as considered in Ref.[\onlinecite{zuev}].
Here, (a) $ S_{xx}^g $, (b) $S_{xx}^d $, (c) $ S_{xx} = S_{xx}^g + S_{xx}^d$.  }
\label{Fig1}
\end{center}
\end{figure}

In Fig 4, we have shown $S_{xx}^g$ as function of $T$ for $B=2.82, 3.88$ and $6.20$ Tesla
(corresponding to three peak values in Fig. 1).
$S_{xx}^g$ increases with increasing $T$, more rapidly at lower $T$. At higher $T$,
the increase is slower and showing nearly
independent behavior for about $T>10$ K. This behavior is similar to the observations 
in conventional 2DEG \cite{flet-1986,kuba,lyo}. The faster increase
of $S_{xx}^g$ with $T$, at low $T$, may be attributed to the increasing number 
of phonons linearly with $T$. For a given magnetic field, maximum momentum transfer 
takes place when $\hbar v_sq \simeq\Gamma$ setting limit on $q$.
As $T$ increases further, the allowed $q$ is limited by the width of LL 
($\Gamma\sim B^{1/2}$) and fewer phonons will exchange momentum. 
In zero magnetic field, such behavior is generally interpreted \cite{but,flet-2003} 
in terms of the dominant phonon wave vector $q_D$ and Fermi wave
vector $2k_F$. At a given $T$,
$S_{xx}^g$  is larger for larger $B$. 
This is consistent with the findings in conventional 2DEG\cite{flet-1986,lyo}.

\begin{figure}[h!]
\begin{center}\leavevmode
\includegraphics[width=100mm,height=60mm]{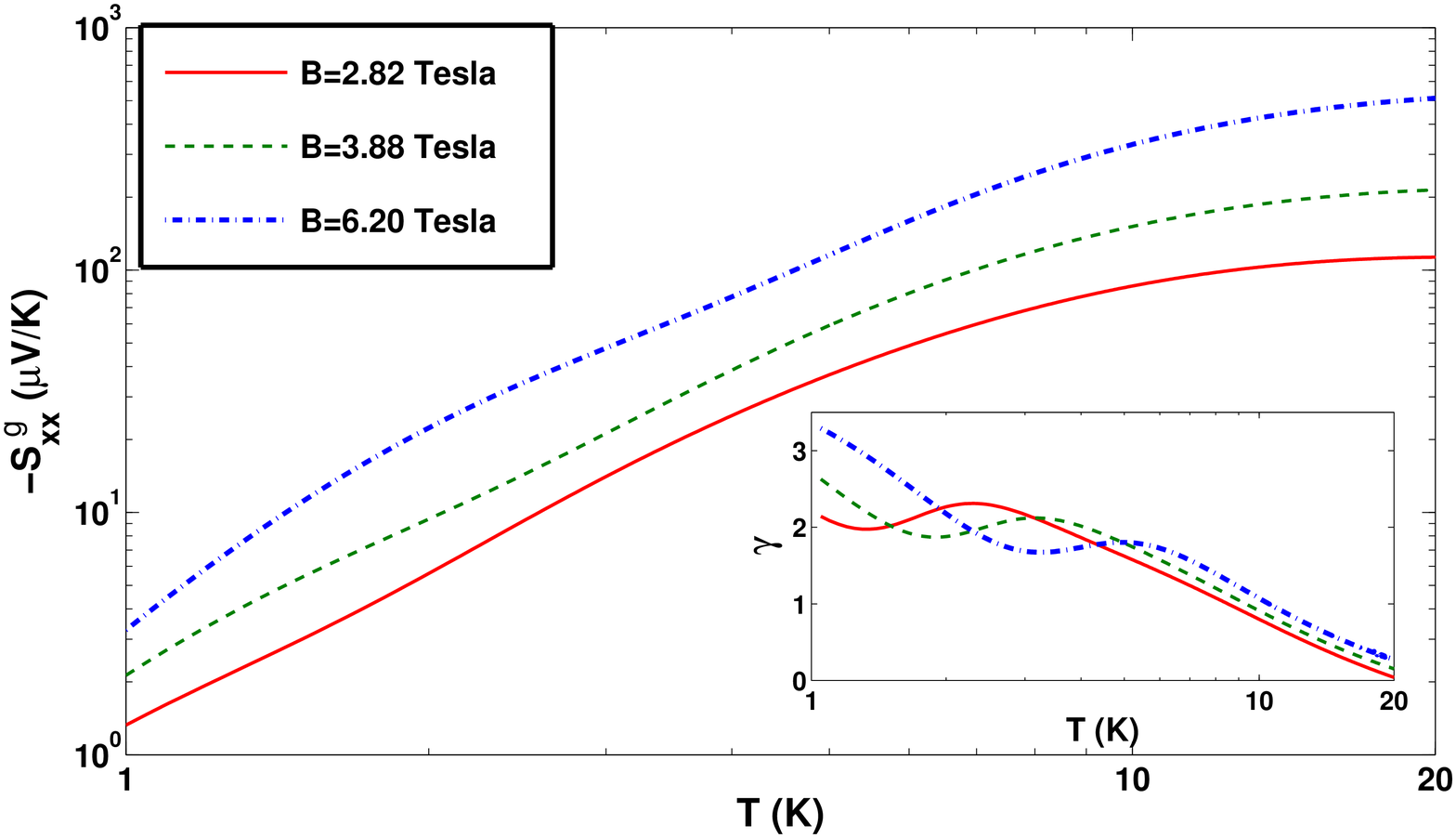}
\caption{(Color online) Temperature dependence of phonon-drag thermopower at
electron density $n_e=10^{16}$ m$^{-2}$. Inset shows variation of the exponent
$\gamma=d\,{\rm ln}S_{xx}^{g}/d\,{\rm ln}T$ with temperature.}
\label{Fig1}
\end{center}
\end{figure}

Inset of Fig. 4, expressing $S_{xx}^g\sim T^{\gamma}$, shows the behavior 
of exponent $\gamma$ as a function of $T$ for different $B$. It is found to 
decrease and tending to zero with increasing $T$. Moreover, $\gamma$ is found to 
be larger for larger $B$.
We observe that for $T$ closer to $1$ K, $\gamma$ is greater than $2$ which is signature of
phonon-drag thermopower.
When $S_{xx}^g$ is calculated as a function of $T$ 
(not shown in the figure ) for different $n_e$ we expect it to increase with 
increasing $T$ but to be smaller for larger $n_e$. These curves are expected to 
show again nearly independent behavior at higher $T$, more so for smaller $n_e$. 
Exponent $\gamma$, in this case $n_e$ dependent, is again expected to decrease 
with increasing $T$ . 

Important point we notice is that $S_{xx}^g$ can be 
tailored to be as large as few mV/K by reducing $n_e$ and working at larger $B$. 
We suggest that the enhanced phonon drag contribution  $S_{xx}^g$ can be achieved
by polishing the  edge of the sample. It is characterized by a specular parameter
$p$ with its value $0 < p < 1$. The perfect reflecting edge gives $p=1$ and very rough edge
corresponds to $p=0$ (diffusive scattering). Besides, the larger samples can be  grown
on piezoelectric substrates. Woszczyna et al\cite{wosz} have shown that the graphene
samples as large as $150\times 30$ $\mu$m$^2$ on GaAs substrate can be prepared.

We would like to point out that the  screening of el-ph coupling in magnetic field is ignored,
although justification is for zero magnetic field case. In conventional 2DEG screening is found
to reduce the phonon drag thermopower significantly both
in zero and quantizing magnetic field\cite{but,flet,flet-2003,tsa}.
However, screening of el-ph interaction in graphene in magnetic field  is yet to be established.
Low temperature experimental $S_{xx}^g$ may throw some light on significance of screening.
One can extract the experimental phonon drag $S_{xx}^g$ from the experimentally measured $S_{xx}$ values
by subtracting diffusion component using generalized Mott formula\cite{zuev}.

\section{summary}
In summary, we have studied phonon-drag thermopower $S_{xx}^g$
in graphene subjected to a transverse magnetic field. Based on a method, 
described in Ref.[\onlinecite{from}], a modified theory is developed to 
calculate $S_{xx}^g$ quantitatively. 
Dependence of $S_{xx}^g$ on magnetic field, electron density, and 
temperature have been studied. With both magnetic
field and density, $S_{xx}^g$ exhibits oscillatory behavior. 
Interestingly, we have found an enhanced 
phonon-drag thermopower with magnitude of the order of few hundreds $\mu$V/K. 
This value is closer to that obtained
in the case of conventional 2DEG at GaAs based semiconductor hetero interface. 
We attribute this enhanced phonon-drag effect is a consequence of taking the 
high value of phonon-mean free path, namely, $\Lambda=10$ $\mu$m.
We, thus, suggest that phonon-drag effect may have significant contribution in larger 
samples of graphene. We have also shown the density dependence of $S_{xx}^g$ for 
parameter values which were taken in Ref.[\onlinecite{zuev}]. The diffusion thermopower has 
also been calculated for the sake of comparison using modified Girvin-Jonson theory. 
Moreover, the temperature dependence 
of $S_{xx}^g$ is also studied and the exponent of this dependence has been extracted.

\section*{Acknowledgement}
SSK would like to thank M. Tsaousidou and TKG 
would like to thank A. Kundu for useful discussions. 

\appendix{}
\section{}
\subsection{Matrix elements of Electron-phonon coupling in a magnetic field}

For a graphene monolayer, lying in $xy$ plane, with a perpendicular magnetic field ${\bf B}=(0,0,B)$, the
eigen functions, for Landau gauge ${\bf A}=(0,Bx,0)$, are given by\cite{matu}

\begin{eqnarray}
\psi_{\alpha}({\bf r})=\frac{e^{ik_yy}}{\sqrt{L_y}}\chi_{n,k_y}(x)
\end{eqnarray}
with
\begin{eqnarray}
 \chi_{n,k_y}(x)=\frac{1}{\sqrt{2}}\left(
\begin{array}{c}
-i\phi_{n-1}(x)\\
 \phi_n(x)
\end{array}\right).
\end{eqnarray}

Here, $\alpha\equiv (n,k_y)$, $n=0,1,2,3,..$ is the Landau level index, $k_y$ is the $y$-component of electron
wave vector, $\phi_n(x)=\sqrt{1/(2^nn!\sqrt{\pi}l_0)} e^{-(x+x_0)^2/(2l_0^2)}
H_n[(x+x_0)/l_0]$ with $x_0=l_0^2k_y$ is the harmonic oscillator wave function.

We assume that at low temperature, for the graphene on the substrate, electrons interact with only in-plane acoustic phonons via
deformation potential coupling. In suspended graphene, there will be flexural modes, whose contribution is
neglected for the graphene on substrate\cite{Oppen}.
The deformation potential coupling is assumed to be only due to longitudinal acoustic phonons.

The most general form of electron-phonon interaction Hamiltonian is
\begin{eqnarray}
H_{ep}({\bf r})=\sum_{{\bf q}s} \Big[V_{{\bf q}s}e^{i{\bf q}\cdot{\bf r}}a_{{\bf q}s}
+V^\dagger_{{\bf q}s}e^{-i{\bf q}\cdot{\bf r}}a^\dagger_{{\bf q}s}\Big],\nonumber
\end{eqnarray}
where $a_{{\bf q}s}(a^\dagger_{{\bf q}s})$ is the phonon annihilation(creation) operator and 
$V_{{\bf q}s}$ is the matrix element of a particular phonon mode $({\bf q},s)$.
For longitudinal acoustic phonon mode corresponding to deformation potential, the form
of $V_{{\bf q}s}$ is given by $V_{q}=D\big[\hbar\omega_q/(2A_0\rho_m v_s^2)\big]^{1/2}$, where $A_0$ is 
the area of graphene sample, $D$ is the deformation
potential coupling constant, and $\rho_m$ is the areal mass density of graphene.
The electron-acoustic phonon matrix element, for the scattering between the states $\alpha \equiv (n,k_y)$
and $\alpha^\prime \equiv(n^\prime,k_y^\prime)$, is given by 
\begin{eqnarray}
C_{\alpha,\alpha^\prime}(q)=\int \psi_{\alpha^\prime}^\dagger({\bf r}) V_{q}e^{i{\bf q}\cdot{\bf r}}\psi_\alpha({\bf r}) d^2r.
\end{eqnarray}

Substituting for $\psi_\alpha({\bf r})$ and $V_{\bf q}$, we get Eq. (\ref{MatEl}) in which the 
integrals are given by 
\begin{eqnarray}
 C_{k_y^\prime, k_y}=\frac{V_{q}}{L_y}\int_0^{L_y} e^{-i(k_y^\prime-k_y-q_y)y}dy=C(q)\delta_{k_y^\prime,k_y+q_y}
\end{eqnarray}

and 
\begin{eqnarray}
J_{n,n^\prime}(u)=\int_{-\infty}^\infty \chi_{n^\prime,k_y^\prime}^\dagger(x)e^{iq_xx}\chi_{n,k_y}(x)dx.
\end{eqnarray}

At low temperature, the acoustic phonon energy is small and cause only intra-Landau level transitions $(n=n^\prime)$.
Inter-Landau level transitions are expected at higher temperatures and in the studies such as magnetophonon resonance
in which optical phonons are involved\cite{Falko}. The matrix element corresponding to intra-Landau level transitions is
found to be 
\begin{eqnarray}
C_{\alpha,\alpha^\prime}(q)=C(q)J_{nn}(u)\delta_{k_y^\prime,k_y+q_y}
\end{eqnarray}
where

\begin{eqnarray}
J_{nn}(u)=\frac{1}{2}e^{iq_xx_0}e^{-\frac{u}{2}}\Big[L_{n-1}(u)+L_n(u)\Big]
\end{eqnarray}
with $u=q^2l_0^2/2$. The equation for $\vert J_{nn}(u)\vert^2$ given in Eq.(\ref{JnnS}) is
similar to the one obtained in Refs.[\onlinecite{Nomura,Kand}].

\end{document}